\documentclass[a4paper,11pt]{article} 
\usepackage{graphicx}
\usepackage{jheppub}
\usepackage{amsmath,amssymb,float,slashed,xcolor,multicol}
\usepackage{enumerate}
\usepackage{caption,subcaption}
\usepackage{float}
\usepackage{microtype}
\usepackage{multirow}
\usepackage[normalem]{ulem}
\usepackage{tablefootnote}
\usepackage[title,titletoc]{appendix}
\title{Charm-quark Yukawa Coupling in $h\rightarrow c\bar{c}\gamma$ at LHC}
\author[a,b]{Tao Han,}
\author[c]{Benjamin Nachman}
\author[a]{and Xing Wang}
\affiliation[a]{Pittsburgh Particle physics, Astrophysics, and Cosmology Center, \\
Department of Physics and Astronomy, University of Pittsburgh, \\
3941 O'Hara St., Pittsburgh, PA 15260, USA}
\affiliation[b]{Department of Physics, Tsinghua University, and Collaborative Innovation Center of Quantum Matter, Beijing, 100086, China}
\affiliation[c]{Physics Division, Lawrence Berkeley National Laboratory, 1 Cyclotron Road, Berkeley, CA 94720, USA}
\emailAdd{than@pitt.edu}
\emailAdd{bpnachman@lbl.gov}
\emailAdd{xiw77@pitt.edu}
\abstract{
It is extremely challenging to probe the charm-quark Yukawa coupling at hadron colliders primarily due to the large Standard Model (SM) background (including $h\to b\bar b$) and the lack of an effective trigger for the signal $h\to c\bar c$. We examine the feasibility of probing this coupling at the LHC via a Higgs radiative decay 
$h\rightarrow c\bar{c}\gamma$.
The existence of an additional photon in the final state may help for the signal identification and background suppression. Adopting a refined triggering strategy and utilizing basic machine learning, we find that a coupling limit of about 8 times the SM value may be reached with $2\sigma$ sensitivity after the High Luminosity LHC (HL-LHC).  
Our result is comparable and complementary to other projections for direct and indirect probes of $h\to c\bar c$ at the HL-LHC. Without a significant change in detector capabilities, there would be no significant improvement for this search from higher energy hadron colliders. 
}
\preprint{
\begin{flushright}
PITT-PACC-1819
\end{flushright}
}

\begin{document}
\maketitle

\section{Introduction}

It is of fundamental importance to establish the pattern of the Higgs boson Yukawa couplings to fermions in order to verify the Standard Model (SM) and seek hints of physics beyond the SM (BSM).  The couplings to third generation fermions have all been observed with over $5\sigma$ significance.  For top quarks, there is a large indirect contribution to the gluon-gluon fusion production mode and the photon-photon decay mode.  However, direct observation is important to ensure there are no BSM quantum corrections to Higgs boson production or decay.  Both the ATLAS and CMS collaborations have recently observed the production of top quark pairs in association to the Higgs boson~\cite{Aaboud:2018urx, Sirunyan:2018hoz} as well as Higgs boson decays to bottom quark pairs~\cite{Aaboud:2018zhk, Sirunyan:2018kst}.  For leptons, the challenging decay channel $h\to \tau^+\tau^-$ reached $5\sigma$ already from the LHC Run 1 data~\cite{Khachatryan:2016vau} and now ATLAS and CMS have both individually observed this decay mode~\cite{Aaboud:2018pen,Sirunyan:2017khh}. 
With the upgrade of the LHC to its high-luminosity phase (HL-LHC), the Higgs coupling measurements to the heaviest generation of fermions will reach an accuracy of about or better than $20\%$ \cite{ATL-PHYS-PUB-2014-016} and will extend to kinematic regions with high transverse momenta of the Higgs boson ($p_T^h$) \cite{Sirunyan:2017dgc,ATLAS-CONF-2018-052}.

Direct observations of the Higgs couplings to the second generation of fermions will be critically important to confirm the pattern of non-universal Yukawa couplings and search for deviations from the SM as predicted in theories with an extended Higgs sector \cite{Branco:2011iw}.
The channel $h\to \mu^+\mu^-$ is the cleanest Higgs signal of all decay modes~\cite{Han:2002gp,Plehn:2001qg}. Even with a branching fraction as small as  $2\times 10^{-4}$, ATLAS and CMS almost have the sensitivity to the SM rate~\cite{Aaboud:2017ojs,Sirunyan:2018hbu} and a measurement with an accuracy of $13\%$ is expected at the HL-LHC \cite{ATL-PHYS-PUB-2018-006}.  
In contrast to $h\to \mu^+\mu^-$, the second generation hadronic decay modes of the Higgs boson are very difficult to distinguish from SM backgrounds, including other Higgs boson decays.  While $b$-jet tagging is a powerful tool for rejecting backgrounds, $c$-jets are harder to distinguish from $b$ and light jets~\cite{Sirunyan:2017ezt,ATLAS:2015ctag,Aaboud:2018fhh} and strange-jets are nearly identical to up- and down-quark jets~\cite{Duarte-Campderros:2018ouv,dshih}.  The $h\to c\bar c$ branching ratio is expected to be about 3\%, so the challenge is background rejection and triggering, not statistics. 

So far, there have been two experimental studies to probe the Higgs-charm Yukawa coupling $(y_c)$.   One approach is to use the clean associated production of the Higgs boson with a vector boson and exploit charm tagging~\cite{Aaboud:2018fhh}.  A key challenge with this method is that the $h\to b\bar b$ contribution is large compared to the $c \bar c$ signal.  An optimistic  projection\footnote{This projection does not account for systematic uncertainties, nor the degradation from the extreme pile-up at the HL-LHC, including the possible increases in lepton trigger thresholds.} for the full HL-LHC dataset suggests that 6 times the SM rate at 95\% confidence level may be achievable~\cite{ATL-PHYS-PUB-2018-016}.  A second approach used exclusive decays of the Higgs into a $J/\psi$ and a photon~\cite{Bodwin:2013gca}.  While this final state can be well-separated from backgrounds in the $J/\psi\rightarrow \mu^+\mu^-$ channel ~\cite{Aaboud:2018txb,Aad:2015sda,Khachatryan:2015lga}, it suffers from a small branching ratio and modeling assumptions to extract the Higgs-charm Yukawa.  In particular, the leading contribution to this process is via the \textit{vector meson dominance} $\gamma^* \to J/\psi$, which is an order of magnitude larger than that involving the charm-quark Yukawa coupling \cite{Bodwin:2013gca, Bodwin:2014bpa,Koenig:2015pha}, leading to a less sensitive upper bound on $y_c$ of about 50 times of the SM prediction at the HL-LHC~\cite{ATL-PHYS-PUB-2015-043}. 

Another recent proposal for probing the Higgs-charm Yukawa coupling is to study the associated production process $gc \to ch$~\cite{Brivio:2015fxa}.  This has the advantage that it is independent of the Higgs decay mode, but suffers from a low rate and significant background.  Other proposals for direct or indirect probes of first- and second-generation quark-Higgs couplings \cite{Soreq:2016rae, Yu:2016rvv, Bishara:2016jga, Cohen:2017rsk} are challenging due to large SM backgrounds and contamination from other Higgs production and decay modes.  A global analysis of Higgs decays can also constrain the charm-Higgs Yukawa coupling, with a projected sensitivity of about 6 times the SM expectation~\cite{Perez:2015aoa,Carpenter:2016mwd}. 

It has been recently pointed out that the radiative decay of the Higgs boson to a pair of charm quarks could be used to constrain the charm-quark Yukawa coupling with the final state $h \to c\bar c \gamma$ \cite{Han:2017yhy}.   The addition of the photon can be helpful for triggering as well as suppressing both non-Higgs and Higgs backgrounds.  In particular, the electromagnetic coupling would disfavor the down-type quarks, especially the flavor-tagged $b\bar b \gamma$ mode.
In this work, we examine the feasibility of a Higgs-charm Yukawa coupling measurement in the $h\rightarrow c\bar{c}\gamma$ channel at the HL-LHC. By proposing an optimal triggering strategy and simulating realistic detector effects, we show that a coupling of about 8 times the SM value may be reached at 95\% confidence after the HL-LHC. This approach is complementary and competitive with other methods.
We also explore the extent to which the energy upgrade of the LHC (HE-LHC) could improve the sensitivity. 

The rest of the paper is organized as follows. In Sec.~\ref{sec:trigger}, we consider the features for the signal and background processes and propose an optimal but realistic trigger. In Sec.~\ref{sec:ana}, we perform detailed analyses, including some basic machine learning, to obtain the optimal signal sensitivity. We extend our analyses to the HE-LHC in Sec.~\ref{sec:HE}. The paper ends with conclusions and outlook in Sec.~\ref{sec:con}.

\section{Trigger Considerations at HL-LHC}
\label{sec:trigger}

We focus on the leading Higgs production channel, gluon fusion, followed by the radiative decay 
\begin{equation}
gg\to h\rightarrow c\bar{c}\gamma .
\end{equation}
The signal is thus characterized by an isolated photon recoiling against two charm-tagged jets with a three-body invariant mass near the Higgs resonance. The energy of the two charm jets will be limited by the  Higgs boson mass, and the photon tends to be soft and collinear with one of the charm quarks.  Due to the large collision rate (40 MHz), enormous inelastic cross-section for $pp\rightarrow$ central activity, and limitations in hardware, most collisions at the LHC are discarded in real time.  The \textit{trigger} system is a key challenge for recording physics processes with relatively soft final states such as $h\rightarrow c\bar{c}\gamma $.  The rest of this section explores the impact of triggering on the $h\to c\bar{c}\gamma $ analysis in the context of the HL-LHC.

\subsection{Signal and background processes}

Tagging jets originating from charm quarks ($c$-tagging) is challenging, but important for suppressing backgrounds originating from light Quantum Chromodynamic (QCD) jets and from $b$-quark jets.  Encouragingly, a recent study from ATLAS \cite{Aaboud:2018fhh} has shown very promising $c$-tagging results.  Based on the ATLAS result, three $c$-tagging working points listed in Table~\ref{tab:c-tag} are studied for the $h\to c\bar c\gamma$ search.\footnote{We choose the $c$-tagging working points aiming at the rejection for the largest background of the QCD light-jets production for given $c$-tagging efficiencies.}

\begin{table}[tb]
	\begin{center}
		\renewcommand{\arraystretch}{1.5}
		\begin{tabular}{|c|c|c|c|}
			\hline
			Operating Point & $\epsilon_c$ & $\epsilon_b$ & $\epsilon_j$ \\
			\hline\hline
			I & 20\%  & 33\%  & 0.13\%  \\
			II & 30\%  & 33\%  & 1\%  \\
			III & 41\%  & 50\%  & 3.3\%  \\
			\hline
		\end{tabular}
	\end{center}
	\caption{Representative operating points for the $c$-tagging efficiency ($\epsilon_c$), the $b$-jet mis-tag rate ($\epsilon_b$), and the light jet mist-tag rate ($\epsilon_j$).}
	\label{tab:c-tag}
\end{table}

One of the dominant backgrounds from the $h\to c\bar c\gamma$ search is QCD di-jet production associated with a photon, where both jets are (mis-)tagged as $c$-jets. Similarly, QCD 3-jet production also contributes to the background if one of the jets is mis-identified as a photon.  In addition to these hard-scatter background processes, one or more of the tagged objects could come from an additional nearly simultaneous $pp$ collision (pile-up). Many sophisticated pile-up mitigation techniques have been proposed~\cite{ATLAS:2017pfq,Berta:2014eza,Bertolini:2014bba,Krohn:2013lba,Cacciari:2014gra,Cacciari:2007fd,CMS:2013wea,Aaboud:2017pou,Aad:2015ina,Martinez:2018fwc} which can significantly reduce the contamination from pile-up both in the trigger and in offline analysis.  However, no method can eliminate all of the pile-up and all methods perform worse (if even applicable) at the trigger level.  Since pile-up conditions will be extreme at the HL-LHC (typically 200 pile-up collisions), their contribution to the event rate must be taken into account.

Current and future upgrades of the ATLAS and CMS trigger systems~\cite{Collaboration:2285584,Collaboration:2283192} will allow for multi-object requirements using offline-like information.   In order to have a high efficiency, (relatively) low rate trigger for $h\to c\bar c\gamma$, we propose a new approach which requires two jets and one photon in the central region with invariant mass near the Higgs resonance.

\subsection{Simulation Setup}

Since the cross section for Higgs bosons is much smaller than for multijet production, the trigger rate is dominated by background.  In order to estimate the trigger rate, the following background processes are simulated using {\tt MG5aMCNLO}~\cite{Alwall:2014hca}, including up to one additional jet matched using the MLM prescription~\cite{Mangano:2002ea}: 

\begin{equation}
pp\rightarrow j\gamma\quad {\rm and}\quad pp\rightarrow jj.
\end{equation}
\begin{sloppypar}
\noindent The parton shower and hadronization are simulated with {\tt PYTHIA6.4.28}~\cite{Sjostrand:2006za}, and a fast detector simulation is implemented using {\tt DELPHES3}~\cite{deFavereau:2013fsa} with the detector card {\tt delphes\_card\_ATLAS\_PilUp.tcl}. Pile-up is modeled  by mixing $\mu=200$ minimum bias events simulated using {\tt PYTHIA} with the hard-scatter processes.
\end{sloppypar}

The ATLAS and CMS trigger systems consist of a hardware trigger (L1) and a software-based high-level trigger (HLT).  While the HLT jet resolution is very similar to offline, at L1, the momentum resolution for jets is much worse than offline due to the coarser detector granularity and reduced information available for the reconstruction algorithms.  The event rate will have a significant contribution from events with low transverse momenta that fluctuate high, since the $p_T$ spectrum is steeply falling.  In order to model the L1 jet resolution, a normal random number is added to each jet energy with a mean of zero and a standard deviation of 13 GeV.  This additional resolution is estimated from the trigger turn-on curves in Ref.~\cite{Aaboud:2016leb} as follows.  Consider a jet trigger that requires a L1 $p_T^\text{L1}>X$ GeV.  The distribution of L1 jet $p_T$ given the offline jet $p_T$ should be approximately Gaussian (ignoring effects from the prior) with a mean $\mu$ and standard deviation $\sigma$.   Suppose that $\Pr(p_T^\text{L1}>X|p_T^\text{offline}=Y)=50\%$.  Since the mean and median of a Gaussian are the same, it must be that for $p_T^\text{offline}=Y$, $\mu=X$.  From Fig.~31a in Ref.~\cite{Aaboud:2016leb}, this procedure gives the relationship $p_T^\text{offline}\sim 2.5\times \mu$ that is nearly independent of $p_T$.  Now, suppose that the same L1 trigger $p_T^\text{L1}>X$ GeV is $99\%$ efficient at $p_T^\text{offline}=Y$ GeV.   This means that the $3\sigma$ tail of the Gaussian with $\mu\sim Y/2.5$ is at $X$.  Therefore, $\sigma\sim (Y/2.5-X)/3$.  Once again using Fig.~31a in Ref.~\cite{Aaboud:2016leb}, this procedure gives $\sigma\sim 5$ GeV, approximately independent of $p_T$.  Translating this 5 GeV back to an offline-scale results in $5\times 2.5\sim 13$ GeV.  Some degradation in this resolution will occur between the LHC and the HL-LHC, but a significant amount of the loss from pile-up will be compensated by gains in performance due to detector upgrades. 

In addition to degrading the resolution of reconstructed jets, pile-up is also a source of jets from additional hard multijet events and random combinations of radiation from multiple soft collisions.  Offline, the most effective method for tagging these pile-up jets is to identify the hard-scatter collision vertex and then record the contribution of momentum from tracks originating from other vertices.  Full-scan tracking and vertexing is not currently available at L1, but both ATLAS and CMS will implement some form of tracking for the HL-LHC~\cite{Collaboration:2285584,Collaboration:2272264,Collaboration:2283193,Collaboration:2296612,Collaboration:2290829,Collaboration:2257755,Collaboration:2285585}.  Using Ref.~\cite{Collaboration:2272264} as an example, we assume a L1 tracking system that has nearly 100\% efficiency for central charged-particle tracks with $p_T>3$ GeV and a $z_0$ resolution of 0.2~cm.  We further assume that some timing information will be available at L1 so that no pile-up tracks with $p_T>3$ GeV enter the analysis.  All of these conditions are optimistic, but are useful when setting a bound on what is achievable with the HL-LHC dataset.  Tracks that can be identified as originating from pile-up are removed before jet clustering so that in a particle-flow-like~\cite{Sirunyan:2017ulk,Aaboud:2017aca} jet reconstruction algorithm, pile-up jets will be reconstructed with less energy than their true energy.  To further suppress pile-up jets, a transverse momentum fraction of tracks within a jet is constructed per jet:

\begin{equation}
r_{c} = \frac{\sum p^{\rm track}_T}{p^{\rm jet}_T},
\end{equation}
where $p^{\rm track}_T$ is the transverse momenta of L1 reconstructable tracks and $p^{\rm jet}_T$ is the transverse momenta of the corresponding jet.   Large values of $r_c$ correspond to more hard-scatter-like jets while low values of $r_c$ are indicative of pile-up jets.  Since the sophisticated pile-up mitigation techniques mentioned earlier can be employed with nearly offline-level performance at the HLT and the pile-up challenge is most severe at L1, the impact of pile-up at the HLT and offline is ignored for the results presented in later sections.

Displaced vertex reconstruction at L1 is likely not possible with high efficiency and so we assume that no explicit $c$-tagging will be possible at L1.  At the HLT, we assume offline-like $c$-tagging.  Flavor tagging does degrade with pile-up, but detector upgrades are expected to compensate for pile-up (Fig. 6 in Ref.~\cite{ATL-PHYS-PUB-2016-026} and Fig. 19a in Ref.~\cite{ATL-PHYS-PUB-2017-013}).

The probability for jets faking photons depends on how well-isolated photon candidates are required to be.  Very stringent isolation requirements result in a purer sample of prompt photons at a cost of signal efficiency while loose requirements result in many fragmentation photons originating from jets.   In our study, we follow the performance evaluation by ATLAS~\cite{ATL-PHYS-PUB-2016-026}, and assume that the fake photon rate would be
\begin{equation}
\epsilon_{j\rightarrow\gamma} = 2.5~(0.7)\times10^{-4},
\end{equation}
for a hard-scatter (pile-up) jet.\footnote{In our simulation, we define hard-scatter jets as jets close to a truth level jet with $\Delta R < 0.3$.} This false positive rate corresponds to an isolation criterion that requires the sum of the transverse energy from the calorimeter within a cone of size $R_c = 0.2$ centered around the photon
candidate $E^{R<R_c}_{T}$ to be
\begin{equation}
E^{R<R_c}_{T} < 6~{\rm GeV}.
\label{eq:iso}
\end{equation}
We further assume the misidentified photons carries 75\% of the jet transverse momenta.

\subsection{Trigger Design}

Currently, the L1 trigger has a maximum rate of 100~kHz, while HLT has a maximum rate of 1~kHz. After the HL-LHC upgrades~\cite{Butler:2055167, CERN-LHCC-2015-020}, the trigger rates at L1 and HLT are expected to be about 1~MHz and 10~kHz, respectively. Therefore, it is vital to make sure the event rates of the processes are within the capacities of both the L1 trigger and the HLT.

For the L1 trigger, we required the two jets and a photon with transverse momenta
\begin{equation}
p_{Tj}>27~{\rm GeV},\;\; p_{T\gamma}>20~{\rm GeV},
\label{eq:cut1}
\end{equation}
and well-separated in the central region
\begin{equation}
|\eta| < 2.5,\;\; {\rm and} \;\;\Delta R > 0.4.
\label{eq:cut2}
\end{equation}
Pile-up jets are rejected by requiring
\begin{equation}
r_c > 0.2.
\label{eq:cut4}
\end{equation}
To suppress the QCD background and put the L1 trigger rate under control, we make use of the fact that the three final state objects come from the Higgs resonance decay. 
Therefore, we also require the invariant mass of the three trigger objects at L1 to be
\begin{equation}
90~{\rm GeV} < M_{jj\gamma} < 160~{\rm GeV}.
\label{eq:cut3}
\end{equation}
As the two jets come from the Higgs decay and do not tend to have rather high transverse momenta, they are often not the two leading jets at L1. Therefore, we require the two candidate jets must be among the 5 hardest jets in each event.

The the corresponding trigger rate is listed in the first row of Table~\ref{tab:Nevents_14}. The trigger rate is calculated using the instantaneous luminosity
\begin{equation}
L=5\times 10^{34}~{\rm cm}^{-2}{\rm s}^{-1} = 5\times 10^{-5}~{\rm fb}^{-1}{\rm s}^{-1} 
\end{equation}
at the HL-LHC~\cite{Apollinari:2017cqg}. We note that the most dominant contribution at L1 comes from the QCD multi-light-jet production with a jet-faked photon. As shown in Table~\ref{tab:Nevents_14}, the trigger proposed above would occupy less than 1\% of the total bandwidth, and thus is plausible to implement as part of the HL-LHC trigger menus of ATLAS and CMS. 

\section{Analyses}
\label{sec:ana}

\subsection{Cut-based Analysis}
\label{sec:cut}
\begin{figure}[tb]
\centering
\begin{subfigure}{0.45\textwidth}
\includegraphics[width=\textwidth]{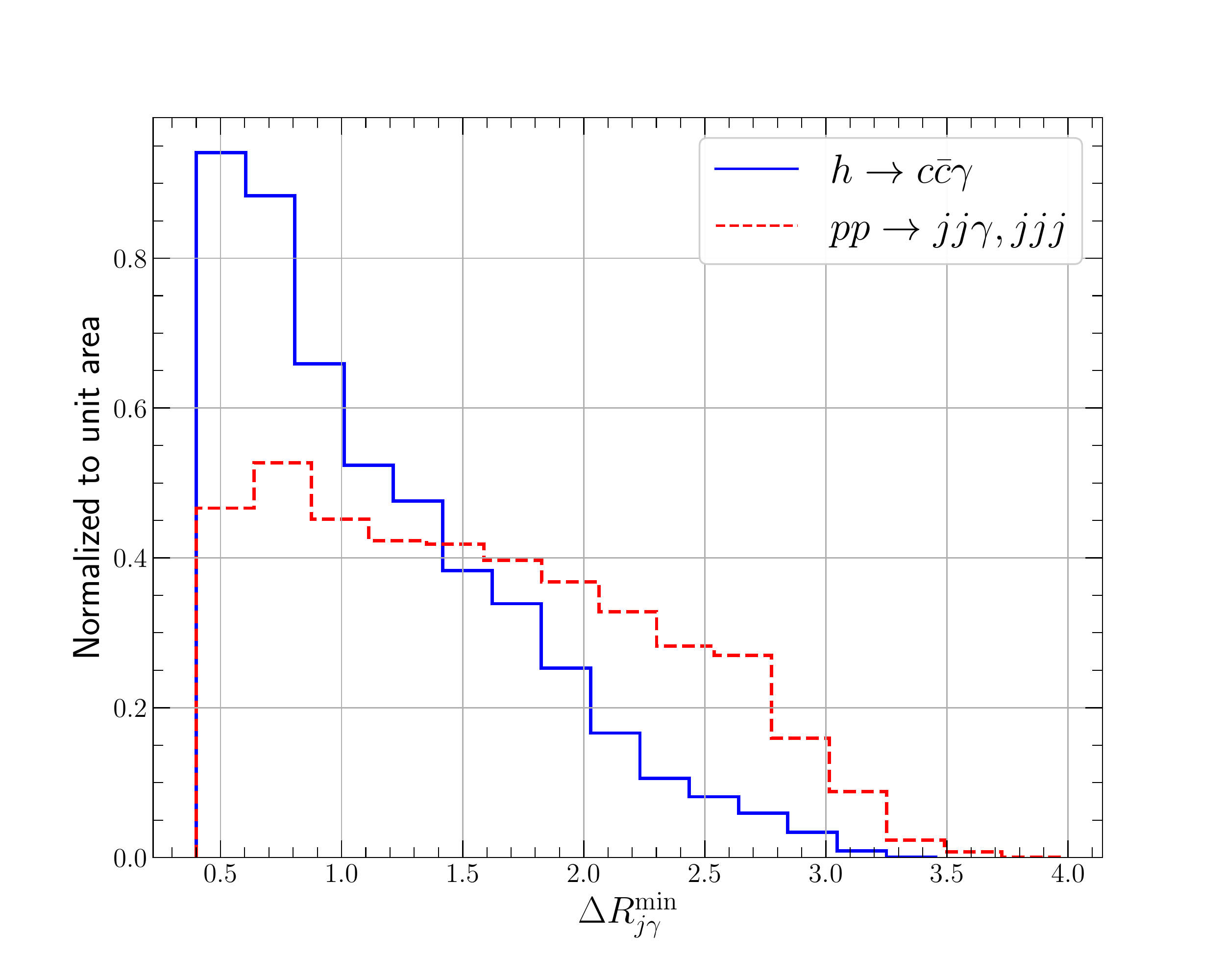}
\caption{}
\label{fig:dRajmin}
\end{subfigure}
\begin{subfigure}{0.45\textwidth}
\includegraphics[width=\textwidth]{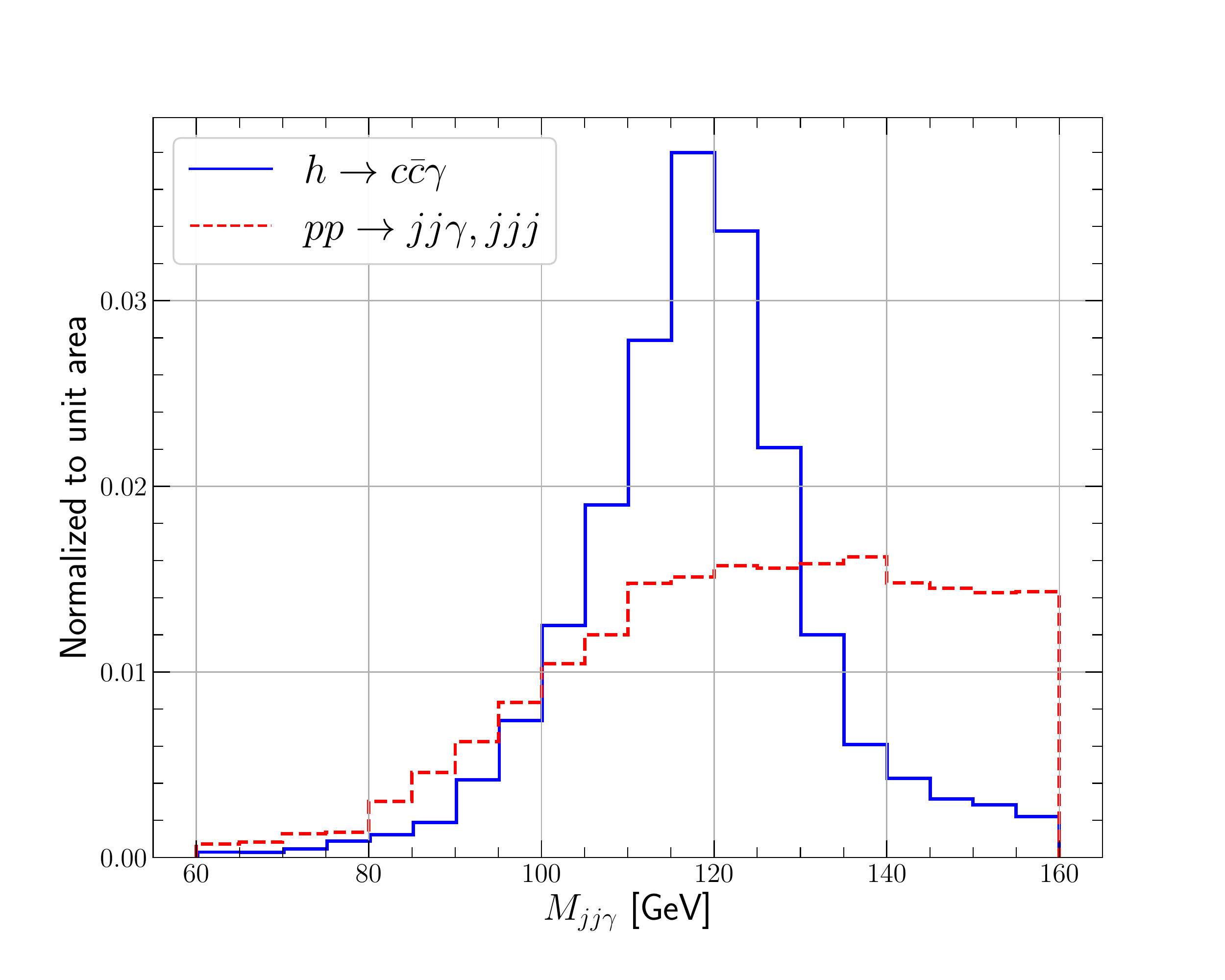}
\caption{}
\label{fig:Mjja}
\end{subfigure}
\caption{Distributions of (a) the smaller value of the separations between the candidate jets and the photon; (b) the three-body invariant mass of the two candidate jets and the photon. Signal (blue solid) and background (red dashed) are both normalized to unit area.}
\end{figure}

To gain physical intuition, we start with a simple analysis that uses only thresholds on various kinematic quantities (``cut-based''). In addition to the trigger requirements as before, we select the signal events with a basic threshold on the leading jet
\begin{equation}
\aligned
&p_{Tj}^{\rm max} > 40~{\rm GeV}.
\endaligned
\label{eq:cut5}
\end{equation}
Figure~\ref{fig:dRajmin} shows the normalized distribution of the smaller value of the separations between photon and jets. As the photon in the signal process comes from final-state radiation, it tends to be close in angle to one of the jets.
Therefore, to optimize the signal significance, we further require the smaller one of the separations between the candidate jets and photon to be
\begin{equation}
\Delta R^{\rm min}_{j\gamma} < 1.8.
\end{equation}
We quantify the sensitivity using a profile likelihood fit to the three-body invariant mass in the range
\begin{equation}
 60~{\rm GeV}  < M_{jj\gamma} < 160~{\rm GeV}, 
 \label{eq:fit_range}
\end{equation}
as shown in Fig.~\ref{fig:Mjja}, with bin widths of 5~GeV, in two event categories. The two categories are defined as having either 1 or 2 of the Higgs candidate jets $c$-tagged.

The expected 95\% ${\rm CL_s}$ \cite{Read:2002hq} upper limit (approximately at a $2\sigma$-level) on the signal strength in the absence of systematic uncertainties is found to be\footnote{In principle, one-loop electroweak processes also contribute to $h\rightarrow c\bar{c}\gamma$, as discussed in \cite{Han:2017yhy}. However, it can be safely neglected given the hypothetically  large Yukawa coupling $y_c$ accessible here and the absence of interference between the QED radiation and EW contributions.}
\begin{equation}
\mu < 106,\;\; 88,\;\; 86,
\end{equation}
for operating points I, II, III with a luminosity of $3~{\rm ab^{-1}}$.
If the BSM physics significantly modifies the charm-Yukawa coupling, which can be parametrized using the $\kappa$-scheme,
\begin{equation}
y_c^{\rm BSM} = \kappa_c \,y_c^{\rm SM},
\end{equation}
then the number of signal events would approximately scale as
\begin{equation}
N_{\rm sig} \simeq \kappa_c^2 N_{\rm sig,QED}^{\rm SM}.
\end{equation}
Then the above upper limit can be translated into
\begin{equation}
\kappa_c = \sqrt{\mu} <10.4,\;\; 9.4,  \;\; 9.3.
\end{equation}
The expected numbers of events and event rates, in the range of $100 < M_{jj\gamma} < 140~{\rm GeV}$, are summarized in Table~\ref{tab:Nevents_14}, for different event categories and $c$-tag working points described in Table~\ref{tab:c-tag}. The third column shows the numbers of events for $h\rightarrow c\bar{c}\gamma$ through QED radiation.
The signal-to-background ratio $S/B$ is between $10^{-5}$ to $10^{-6}$.  As the background is dominated by QCD multi-jet processes, it is likely that the background would be estimated using data-driven techniques.  The resulting systematic uncertainties may not be small, but would likely be comparable to or smaller than the large relative statistical uncertainty on the signal. 
We also note that, although we aimed to optimize the light-jet rejection, the yields of the background process $h\rightarrow b\bar{b}\gamma$ due to mis-tagging is about $1.5-3$ times larger than those of $h\rightarrow c\bar{c}\gamma$ for different $c$-tagging working points, comparable to the previous studies.\footnote{For reference, the background rates for $h\rightarrow b\bar b$ in the $Vh(\to c\bar c)$ searches presented in Ref.~\cite{Aaboud:2018fhh} and Ref.~\cite{ATL-PHYS-PUB-2018-016} are $5-10$ and $1-5$ times higher than the signal $h\rightarrow c\bar c$, respectively, where different $c$-tagging working points are used.}

\begin{table}[tb]
	\begin{center}
		\renewcommand{\arraystretch}{1.5}
       \begin{tabular}{|c|c|c|c||c||c|}
           \hline
                   &            Working        &              Signal  &          Background  &          Background    &      $S/\sqrt{S+B}$  \\
                   &              Point        &               (QED)  &             events         &          event rate [Hz]  &         $[10^{-2}]$  \\
           \hline\hline
           Level-1 (L1) & No Tag    &   -  &   -  & $ 9.55\times 10^{3}$ & - \\
           \hline
                   & I         & $ 269$ & $ 3.37\times 10^{8}$ & $              5.62$ & $              1.47$ \\
           1 $c$-tag & II        & $ 349$ & $ 5.18\times 10^{8}$ & $              8.63$ & $              1.54$ \\
                   & III       & $ 401$ & $ 8.83\times 10^{8}$ & $              14.7$ & $              1.35$ \\
           \hline
                   & I         & $  29$ & $ 1.14\times 10^{7}$ & $             0.191$ & $             0.878$ \\
           2 $c$-tags& II        & $  66$ & $ 2.23\times 10^{7}$ & $             0.371$ & $              1.42$ \\
                   & III       & $ 126$ & $ 5.79\times 10^{7}$ & $             0.966$ & $              1.66$ \\
           \hline
       \end{tabular}
	\end{center}
	\caption{Expected numbers of events of the signal and background, and event rates, in the range of $100 < M_{jj\gamma} < 140~{\rm GeV}$ at the HL-LHC with $\mathcal{L} = 3~{\rm ab}^{-1}$. The first row gives the event rate at L1, with only the requirements in Sec.~\ref{sec:trigger} applied.  Systematic uncertainties are not accounted for in the significance calculation in the last column.}
	\label{tab:Nevents_14}
\end{table}

\subsection{Machine Learning Analysis}

\begin{figure}[tb]
\centering
\begin{subfigure}{0.45\textwidth}
\includegraphics[width=\textwidth]{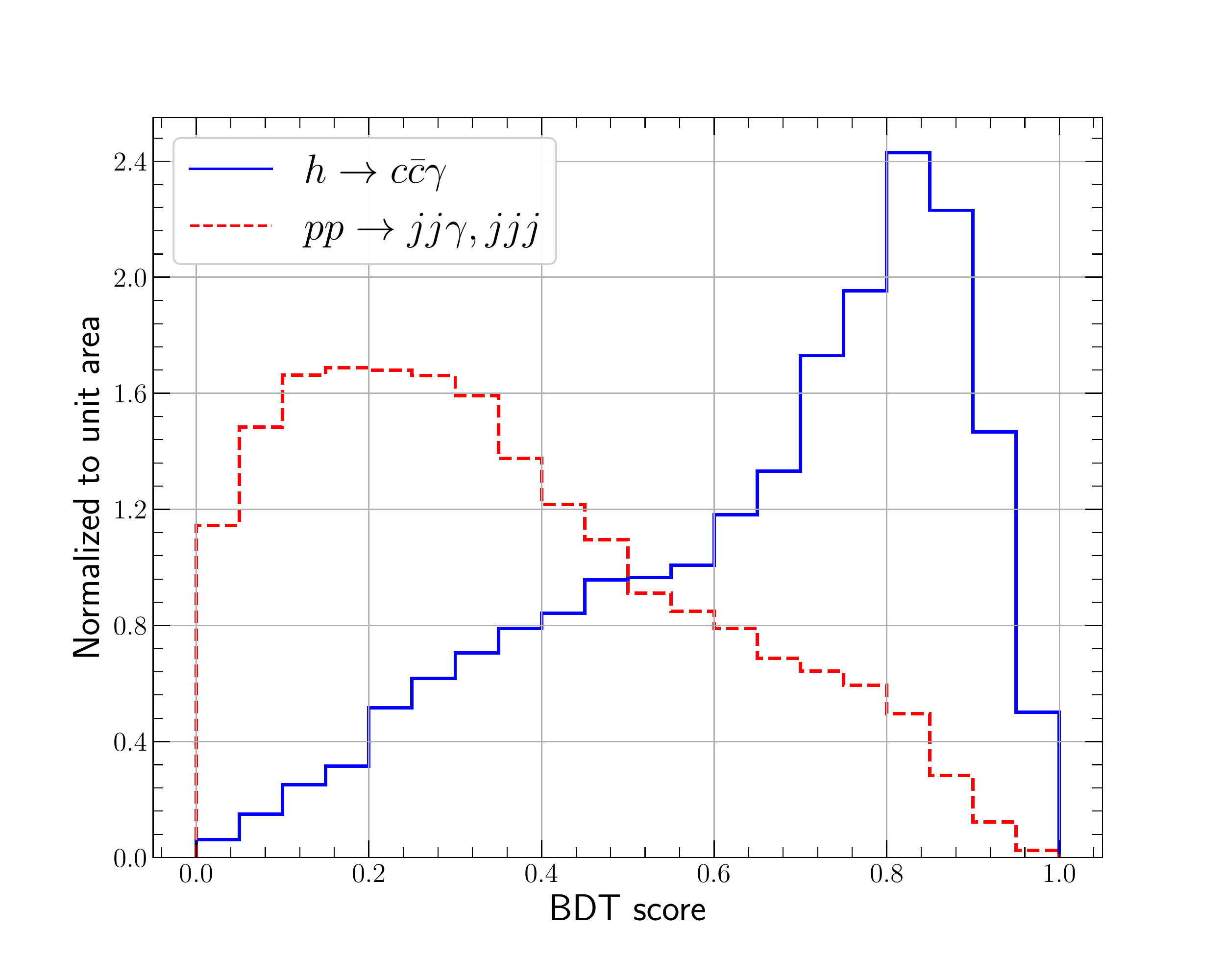}
\caption{}
\label{fig:BDT_score}
\end{subfigure}
\begin{subfigure}{0.45\textwidth}
\includegraphics[width=\textwidth]{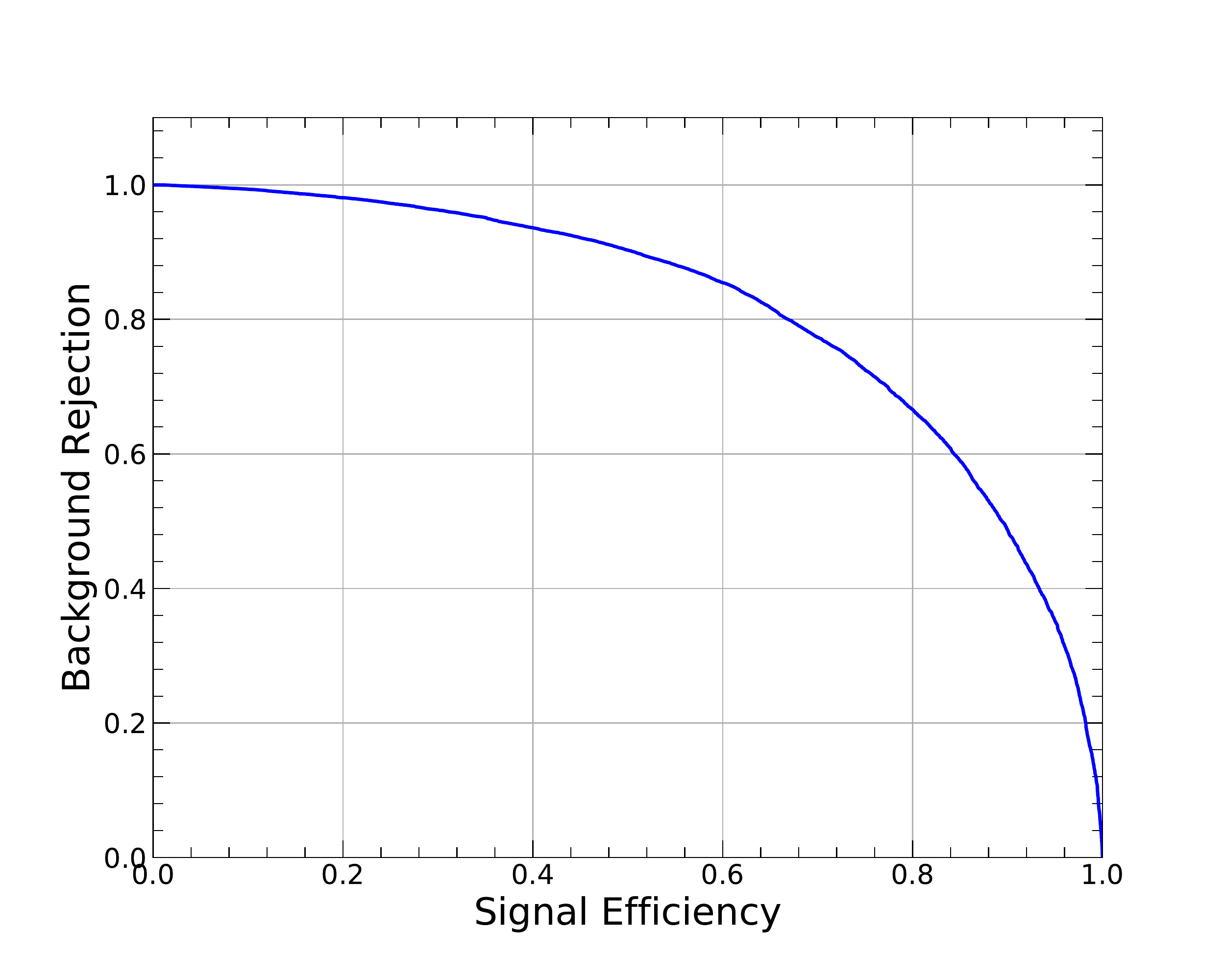}
\caption{}
\label{fig:roc}
\end{subfigure}
\caption{(a) The distribution of the BDT score normalized to unit area. (b) Receiver operating characteristic (ROC) curve of the BDT.} 
\label{fig:BDT}
\end{figure}

In order to study the benefit from a more complex analysis approach, a boosted decision tree (BDT) is trained to distinguish the Higgs signal from the multi-jet background.  The BDT is trained using  {\tt XGBoost}~\cite{Chen_2016} with 5-fold cross-validation.  The following 13 input features are used for training:

\begin{equation}
M_{j\gamma}^{\rm max},\ M_{j\gamma}^{\rm min},\ M_{jj},\ p_{T\gamma},\ p_{Tj}^{\rm max},\ p_{Tj}^{\rm min},\ \eta_\gamma,\ \eta_j^{\rm max},\ \eta_j^{\rm min},\ \Delta R_{j\gamma}^{\rm max},\ \Delta R_{j\gamma}^{\rm min},\ \Delta R_{jj},\ p_{Tjj\gamma} .
\end{equation}
Even though $M_{jj\gamma}$ is the most important feature, it is not explicitly provided to the BDT in order to minimize the bias to the distribution used for the profile likelihood fit in the range of Eq.~(\ref{eq:fit_range}) for extracting the expected upper limit.\footnote{There are many methods for performing this decorrelation using more explicit and even automated methods~\cite{Louppe:2016ylz,Shimmin:2017mfk,Aguilar-Saavedra:2017rzt,Stevens:2013dya,Dolen:2016kst,Moult:2017okx,ATL-PHYS-PUB-2018-014}.}   The distribution of the BDT output on signal and background along with a receiver operator characteristic (ROC) curve are shown in Fig.~\ref{fig:BDT}. The two most important features used by the BDT are $p_{Tj}^{\rm max}$ and $\Delta R_{j\gamma}^{\rm min}$, which are also the features used to form the simple event selection in the previous section.

Using a selection based on the BDT, the expected 95\% ${\rm CL_s}$ upper limit on the signal strength in the absence of systematic uncertainties is found to be
\begin{equation}
\mu < 91,\;\; 77,\;\; 75, \quad \Rightarrow\quad \kappa_c <9.6,\;\; 8.8,  \;\; 8.6.
\end{equation}
for operating points I, II, III with a luminosity of $3~{\rm ab^{-1}}$.  This is a modest improvement over the cut-based result by about 10\%.  Further gains using multivariate approaches may be possible, but will likely require advances in photon, pile-up, and $c$-tagging using low-level information.  The distribution of $M_{jj\gamma}$ already captures most of the information available for separating signal and background given that the correct objects are identified.

\section{HE-LHC Projection}
\label{sec:HE}

Given the recent proposal of an energy upgrade (HE-LHC) operating at 
$\sqrt s =27$~TeV~\cite{1742-6596-1067-2-022009} after the high-luminosity phase (HL-LHC), it would be informative to estimate the potential reach for the radiative decay $h\to c\bar c\gamma$. However, it would be a non-trivial job to do so without knowing the high pile-up and the detector performance under the new conditions. As such, for the purpose of illustration, we can only give a crude projection by assuming a similar environment as in the above studies for HL-LHC.
We consider the option with luminosity $\mathcal{L} = 3~{\rm ab}^{-1}$ and the same pile-up $\mu=200$. 
We also assume the same L1 trigger rate.

To compensate the larger and harder background at 27~TeV, we raise the trigger threshold to 
\begin{equation}
p_{Tj}>40~{\rm GeV},\;\; p_{T\gamma}>23~{\rm GeV},
\end{equation}
in order to maintain the same L1 event rate.  As future experiments would come with significant improvements, we relax the isolation cut in Eq.~(\ref{eq:iso}) and conservatively assume that the same fake photon rate can be achieved while the photon isolation efficiency remains unchanged for photon with $p_T>20$~GeV. 

The expected 95\% ${\rm CL_s}$ upper limit on the signal strength via cut-based analysis is found to be
\begin{equation}
\mu < 98,\;\; 82,\;\; 81, \quad \Rightarrow\quad \kappa_c <9.9,\;\; 9.0,  \;\; 9.0.
\end{equation}
for operating points I, II, III with a luminosity of $3~{\rm ab^{-1}}$, in comparison with the expected 95\% ${\rm CL_s}$ upper limit on the signal strength via BDT-based analysis to be
\begin{equation}
\mu < 89,\;\; 71,\;\; 70, \quad\Rightarrow\quad \kappa_c <9.4,\;\; 8.4,  \;\; 8.3.
\end{equation}
We thus do not find significant improvement for probing the charm-quark Yukawa coupling at the high energy upgrade of the LHC, since the sensitivity is mostly limited by the L1 rate which in this work is assumed to stay the same at the HE-LHC.
We would like to reiterate that the estimated projection here should be considered in the context of our assumptions since the results sensitively depend on the unknown pile-up and the detector performance.  Given our assumptions, there is room for potential improvements should the HE-LHC experiments be constructed.

\section{Conclusion}
\label{sec:con}

While it is of fundamental importance to probe the charm-quark Yukawa coupling, it is extremely challenging at hadron colliders primarily due to the SM background and the lack of an effective trigger for the signal $h\to c\bar c$. We pointed out that the branching fraction for the Higgs radiative decay $h\rightarrow c\bar{c}\gamma$ is about $4\times 10^{-4}$ and thus would yield a large number of events at the HL-LHC. 
The existence of an additional photon in the final state may help for the signal identification and background suppression. For instance, the electromagnetic coupling would disfavor the down-type quarks, especially the flavor-tagged $b\bar b \gamma$ mode. 
We thus proposed to take advantage of the radiative decay and examine the feasibility of probing the charm-quark Yukawa coupling. 
We proposed a refined triggering strategy that also included many event features combined with a boosted decision tree.  Our results can be summarized as follows. 
\begin{itemize}
\item 
A traditional cut-based analysis for identifying the signal $h\to c\bar c \gamma$ yields the sensitivity for a coupling of about 9 times of the SM value at $2\sigma$ level at the HL-LHC. 
\item
A boosted decision tree improves the sensitivity by about 10\%, reaching 
a coupling limit of about 8 times the SM value at $2\sigma$. 
\item
As a crude estimate for the sensitivity reach at the HE-LHC, assuming the same pile-up and L1 trigger rate, we found no significant improvement over the results of HL-LHC. There is room for improvement given the assumptions about the HE-LHC experiments and running conditions.
 \end{itemize}
Our results with semi-realistic simulations are comparable to the other related studies \cite{Brivio:2015fxa, Perez:2015aoa,Carpenter:2016mwd, Soreq:2016rae, Yu:2016rvv, Bishara:2016jga, Cohen:2017rsk} and better than the $h\to J/\psi+\gamma$ channel \cite{ATL-PHYS-PUB-2015-043} in constraining the charm-Yukawa coupling. Although slightly weaker than the sensitivity from the ATLAS direct search of about 3 times of the SM value \cite{ATL-PHYS-PUB-2018-016}, there are uncertainties in both analyses due to missing effects in one or the other and so more detailed experimental studies would be required to know which method will achieve the best precision. 
Multiple complementary approaches are needed to improve the sensitivity to test the SM prediction.

We close by making a few remarks on the possible future improvement. 
Since one of the limiting factors is the huge L1 event rate from QCD multi-jets background, a better photon identification would significantly improve the results.   Furthermore, improved $c$-tagging would also enhance the sensitivity and the machine learning techniques would be more beneficial there.  Finally, extending the analysis to other Higgs production modes and different kinematic regimes may help with the trigger challenge.

\section*{Acknowledgments}
TH and XW were supported in part by the U.S.~Department of Energy under contract DE-FG02-95ER40896 and in part by the PITT PACC. BN was supported by the U.S.~Department of Energy under contract DE-AC02-05CH11231. 
XW was also supported by an Andrew Mellon Fellowship from the Dietrich School of Arts and Sciences, the University of Pittsburgh. TH also would like to acknowledge the hospitality for the Aspen Center for Physics, which is supported by National Science Foundation grant PHY-1607611. 

\newpage

\bibliographystyle{JHEP}
\bibliography{refs} 

\providecommand{\href}[2]{#2}\begingroup\raggedright\begin{thebibliography}{10}

\bibitem{Aaboud:2018urx}
{\scshape ATLAS} collaboration, M.~Aaboud et~al., \emph{{Observation of Higgs
  boson production in association with a top quark pair at the LHC with the
  ATLAS detector}},
  \href{http://dx.doi.org/10.1016/j.physletb.2018.07.035}{\emph{Phys. Lett.}
  {\bfseries B784} (2018) 173--191},
  [\href{https://arxiv.org/abs/1806.00425}{{\ttfamily 1806.00425}}].

\bibitem{Sirunyan:2018hoz}
{\scshape CMS} collaboration, A.~M. Sirunyan et~al., \emph{{Observation of
  $\mathrm{t\overline{t}}$H production}},
  \href{http://dx.doi.org/10.1103/PhysRevLett.120.231801,
  10.1130/PhysRevLett.120.231801}{\emph{Phys. Rev. Lett.} {\bfseries 120}
  (2018) 231801}, [\href{https://arxiv.org/abs/1804.02610}{{\ttfamily
  1804.02610}}].

\bibitem{Aaboud:2018zhk}
{\scshape ATLAS} collaboration, M.~Aaboud et~al., \emph{{Observation of $H
  \rightarrow b\bar{b}$ decays and $VH$ production with the ATLAS detector}},
  \href{http://dx.doi.org/10.1016/j.physletb.2018.09.013}{\emph{Phys. Lett.}
  {\bfseries B786} (2018) 59--86},
  [\href{https://arxiv.org/abs/1808.08238}{{\ttfamily 1808.08238}}].

\bibitem{Sirunyan:2018kst}
{\scshape CMS} collaboration, A.~M. Sirunyan et~al., \emph{{Observation of
  Higgs boson decay to bottom quarks}},
  \href{http://dx.doi.org/10.1103/PhysRevLett.121.121801}{\emph{Phys. Rev.
  Lett.} {\bfseries 121} (2018) 121801},
  [\href{https://arxiv.org/abs/1808.08242}{{\ttfamily 1808.08242}}].

\bibitem{Khachatryan:2016vau}
{\scshape ATLAS, CMS} collaboration, G.~Aad et~al., \emph{{Measurements of the
  Higgs boson production and decay rates and constraints on its couplings from
  a combined ATLAS and CMS analysis of the LHC pp collision data at $
  \sqrt{s}=7 $ and 8 TeV}},
  \href{http://dx.doi.org/10.1007/JHEP08(2016)045}{\emph{JHEP} {\bfseries 08}
  (2016) 045}, [\href{https://arxiv.org/abs/1606.02266}{{\ttfamily
  1606.02266}}].

\bibitem{Aaboud:2018pen}
{\scshape ATLAS} collaboration, M.~Aaboud et~al., \emph{{Cross-section
  measurements of the Higgs boson decaying into a pair of $\tau$-leptons in
  proton-proton collisions at $\sqrt{s}=13$ TeV with the ATLAS detector}},
  \href{https://arxiv.org/abs/1811.08856}{{\ttfamily 1811.08856}}.

\bibitem{Sirunyan:2017khh}
{\scshape CMS} collaboration, A.~M. Sirunyan et~al., \emph{{Observation of the
  Higgs boson decay to a pair of $\tau$ leptons with the CMS detector}},
  \href{http://dx.doi.org/10.1016/j.physletb.2018.02.004}{\emph{Phys. Lett.}
  {\bfseries B779} (2018) 283--316},
  [\href{https://arxiv.org/abs/1708.00373}{{\ttfamily 1708.00373}}].

\bibitem{ATL-PHYS-PUB-2014-016}
\emph{{Projections for measurements of Higgs boson signal strengths and
  coupling parameters with the ATLAS detector at a HL-LHC}},  Tech. Rep.
  ATL-PHYS-PUB-2014-016, CERN, Geneva, Oct, 2014.

\bibitem{Sirunyan:2017dgc}
{\scshape CMS} collaboration, A.~M. Sirunyan et~al., \emph{{Inclusive search
  for a highly boosted Higgs boson decaying to a bottom quark-antiquark pair}},
  \href{http://dx.doi.org/10.1103/PhysRevLett.120.071802}{\emph{Phys. Rev.
  Lett.} {\bfseries 120} (2018) 071802},
  [\href{https://arxiv.org/abs/1709.05543}{{\ttfamily 1709.05543}}].

\bibitem{ATLAS-CONF-2018-052}
{\scshape ATLAS Collaboration} collaboration, \emph{{Search for boosted
  resonances decaying to two b-quarks and produced in association with a jet at
  $\sqrt{s}=13$ TeV with the ATLAS detector}},  Tech. Rep. ATLAS-CONF-2018-052,
  CERN, Geneva, Nov, 2018.

\bibitem{Branco:2011iw}
G.~C. Branco, P.~M. Ferreira, L.~Lavoura, M.~N. Rebelo, M.~Sher and J.~P.
  Silva, \emph{{Theory and phenomenology of two-Higgs-doublet models}},
  \href{http://dx.doi.org/10.1016/j.physrep.2012.02.002}{\emph{Phys. Rept.}
  {\bfseries 516} (2012) 1--102},
  [\href{https://arxiv.org/abs/1106.0034}{{\ttfamily 1106.0034}}].

\bibitem{Han:2002gp}
T.~Han and B.~McElrath, \emph{{h to $\mu^+\mu^-$ via gluon fusion at the LHC}},
  \href{http://dx.doi.org/10.1016/S0370-2693(02)01208-X}{\emph{Phys. Lett.}
  {\bfseries B528} (2002) 81--85},
  [\href{https://arxiv.org/abs/hep-ph/0201023}{{\ttfamily hep-ph/0201023}}].

\bibitem{Plehn:2001qg}
T.~Plehn and D.~L. Rainwater, \emph{{Higgs Decays to Muons in Weak Boson
  Fusion}}, \href{http://dx.doi.org/10.1016/S0370-2693(01)01157-1}{\emph{Phys.
  Lett.} {\bfseries B520} (2001) 108--114},
  [\href{https://arxiv.org/abs/hep-ph/0107180}{{\ttfamily hep-ph/0107180}}].

\bibitem{Aaboud:2017ojs}
{\scshape ATLAS} collaboration, M.~Aaboud et~al., \emph{{Search for the dimuon
  decay of the Higgs boson in $pp$ collisions at $\sqrt{s}$ = 13 TeV with the
  ATLAS detector}},
  \href{http://dx.doi.org/10.1103/PhysRevLett.119.051802}{\emph{Phys. Rev.
  Lett.} {\bfseries 119} (2017) 051802},
  [\href{https://arxiv.org/abs/1705.04582}{{\ttfamily 1705.04582}}].

\bibitem{Sirunyan:2018hbu}
{\scshape CMS} collaboration, A.~M. Sirunyan et~al., \emph{{Search for the
  Higgs boson decaying to two muons in proton-proton collisions at $\sqrt{s}=$
  13 TeV}}, {\emph{Submitted to: Phys. Rev. Lett.} (2018) },
  [\href{https://arxiv.org/abs/1807.06325}{{\ttfamily 1807.06325}}].

\bibitem{ATL-PHYS-PUB-2018-006}
{\scshape ATLAS Collaboration} collaboration, \emph{{Prospects for the
  measurement of the rare Higgs boson decay $H\to\mu\mu$ with 3000 fb$^{-1}$ of
  $pp$ collisions collected at $\sqrt{s} = 14$ TeV by the ATLAS experiment}},
  Tech. Rep. ATL-PHYS-PUB-2018-006, CERN, Geneva, May, 2018.

\bibitem{Sirunyan:2017ezt}
{\scshape CMS} collaboration, A.~M. Sirunyan et~al., \emph{{Identification of
  heavy-flavour jets with the CMS detector in pp collisions at 13 TeV}},
  \href{http://dx.doi.org/10.1088/1748-0221/13/05/P05011}{\emph{JINST}
  {\bfseries 13} (2018) P05011},
  [\href{https://arxiv.org/abs/1712.07158}{{\ttfamily 1712.07158}}].

\bibitem{ATLAS:2015ctag}
\emph{{Performance and Calibration of the JetFitterCharm Algorithm for c-Jet
  Identification}},  Tech. Rep. ATL-PHYS-PUB-2015-001, CERN, Geneva, Jan, 2015.

\bibitem{Aaboud:2018fhh}
{\scshape ATLAS} collaboration, M.~Aaboud et~al., \emph{{Search for the Decay
  of the Higgs Boson to Charm Quarks with the ATLAS Experiment}},
  \href{http://dx.doi.org/10.1103/PhysRevLett.120.211802}{\emph{Phys. Rev.
  Lett.} {\bfseries 120} (2018) 211802},
  [\href{https://arxiv.org/abs/1802.04329}{{\ttfamily 1802.04329}}].

\bibitem{Duarte-Campderros:2018ouv}
J.~Duarte-Campderros, G.~Perez, M.~Schlaffer and A.~Soffer, \emph{{Probing the
  strange Higgs coupling at lepton colliders using light-jet flavor tagging}},
  \href{https://arxiv.org/abs/1811.09636}{{\ttfamily 1811.09636}}.

\bibitem{dshih}
D.~S. Y.~Nakai and S.~Thomas, \emph{{Deep Learning Strange Jets}},
  {\emph{https://indico.cern.ch/event/745718/contributions/3174401/attachments/1753217/2841520/StrangeTagV5.pdf}
  (2018) }.

\bibitem{ATL-PHYS-PUB-2018-016}
{\scshape ATLAS Collaboration} collaboration, \emph{{Prospects for
  $H\rightarrow c\bar c$ using Charm Tagging with the ATLAS Experiment at the
  HL-LHC}},  Tech. Rep. ATL-PHYS-PUB-2018-016, CERN, Geneva, Aug, 2018.

\bibitem{Bodwin:2013gca}
G.~T. Bodwin, F.~Petriello, S.~Stoynev and M.~Velasco, \emph{{Higgs boson
  decays to quarkonia and the $H\bar{c}c$ coupling}},
  \href{http://dx.doi.org/10.1103/PhysRevD.88.053003}{\emph{Phys. Rev.}
  {\bfseries D88} (2013) 053003},
  [\href{https://arxiv.org/abs/1306.5770}{{\ttfamily 1306.5770}}].

\bibitem{Aaboud:2018txb}
{\scshape ATLAS} collaboration, M.~Aaboud et~al., \emph{{Searches for exclusive
  Higgs and $Z$ boson decays into $J/\psi\gamma$, $\psi(2S)\gamma$, and
  $\Upsilon(nS)\gamma$ at $\sqrt{s}=13$ TeV with the ATLAS detector}},
  \href{http://dx.doi.org/10.1016/j.physletb.2018.09.024}{\emph{Phys. Lett.}
  {\bfseries B786} (2018) 134--155},
  [\href{https://arxiv.org/abs/1807.00802}{{\ttfamily 1807.00802}}].

\bibitem{Aad:2015sda}
{\scshape ATLAS} collaboration, G.~Aad et~al., \emph{{Search for Higgs and Z
  Boson Decays to J/$\psi\gamma$ and $\Upsilon(nS)\gamma$ with the ATLAS
  Detector}},
  \href{http://dx.doi.org/10.1103/PhysRevLett.114.121801}{\emph{Phys. Rev.
  Lett.} {\bfseries 114} (2015) 121801},
  [\href{https://arxiv.org/abs/1501.03276}{{\ttfamily 1501.03276}}].

\bibitem{Khachatryan:2015lga}
{\scshape CMS} collaboration, V.~Khachatryan et~al., \emph{{Search for a Higgs
  boson decaying into $\gamma^* \gamma \to \ell \ell \gamma$ with low dilepton
  mass in pp collisions at $\sqrt s = $ 8 TeV}},
  \href{http://dx.doi.org/10.1016/j.physletb.2015.12.039}{\emph{Phys. Lett.}
  {\bfseries B753} (2016) 341--362},
  [\href{https://arxiv.org/abs/1507.03031}{{\ttfamily 1507.03031}}].

\bibitem{Bodwin:2014bpa}
G.~T. Bodwin, H.~S. Chung, J.-H. Ee, J.~Lee and F.~Petriello,
  \emph{{Relativistic corrections to Higgs boson decays to quarkonia}},
  \href{http://dx.doi.org/10.1103/PhysRevD.90.113010}{\emph{Phys. Rev.}
  {\bfseries D90} (2014) 113010},
  [\href{https://arxiv.org/abs/1407.6695}{{\ttfamily 1407.6695}}].

\bibitem{Koenig:2015pha}
M.~K{\"o}nig and M.~Neubert, \emph{{Exclusive Radiative Higgs Decays as Probes
  of Light-Quark Yukawa Couplings}},
  \href{http://dx.doi.org/10.1007/JHEP08(2015)012}{\emph{JHEP} {\bfseries 08}
  (2015) 012}, [\href{https://arxiv.org/abs/1505.03870}{{\ttfamily
  1505.03870}}].

\bibitem{ATL-PHYS-PUB-2015-043}
\emph{{Search for the Standard Model Higgs and Z Boson decays to
  $J/\psi\,\gamma$: HL-LHC projections}},  Tech. Rep. ATL-PHYS-PUB-2015-043,
  CERN, Geneva, Sep, 2015.

\bibitem{Brivio:2015fxa}
I.~Brivio, F.~Goertz and G.~Isidori, \emph{{Probing the Charm Quark Yukawa
  Coupling in Higgs+Charm Production}},
  \href{http://dx.doi.org/10.1103/PhysRevLett.115.211801}{\emph{Phys. Rev.
  Lett.} {\bfseries 115} (2015) 211801},
  [\href{https://arxiv.org/abs/1507.02916}{{\ttfamily 1507.02916}}].

\bibitem{Soreq:2016rae}
Y.~Soreq, H.~X. Zhu and J.~Zupan, \emph{{Light quark Yukawa couplings from
  Higgs kinematics}},
  \href{http://dx.doi.org/10.1007/JHEP12(2016)045}{\emph{JHEP} {\bfseries 12}
  (2016) 045}, [\href{https://arxiv.org/abs/1606.09621}{{\ttfamily
  1606.09621}}].

\bibitem{Yu:2016rvv}
F.~Yu, \emph{{Phenomenology of Enhanced Light Quark Yukawa Couplings and the
  $W^\pm h$ Charge Asymmetry}},
  \href{http://dx.doi.org/10.1007/JHEP02(2017)083}{\emph{JHEP} {\bfseries 02}
  (2017) 083}, [\href{https://arxiv.org/abs/1609.06592}{{\ttfamily
  1609.06592}}].

\bibitem{Bishara:2016jga}
F.~Bishara, U.~Haisch, P.~F. Monni and E.~Re, \emph{{Constraining Light-Quark
  Yukawa Couplings from Higgs Distributions}},
  \href{https://arxiv.org/abs/1606.09253}{{\ttfamily 1606.09253}}.

\bibitem{Cohen:2017rsk}
J.~Cohen, S.~Bar-Shalom, G.~Eilam and A.~Soni, \emph{{Light-quarks Yukawa
  couplings and new physics in exclusive high- $p_T$ Higgs boson+jet and Higgs
  boson + b -jet events}},
  \href{http://dx.doi.org/10.1103/PhysRevD.97.055014}{\emph{Phys. Rev.}
  {\bfseries D97} (2018) 055014},
  [\href{https://arxiv.org/abs/1705.09295}{{\ttfamily 1705.09295}}].

\bibitem{Perez:2015aoa}
G.~Perez, Y.~Soreq, E.~Stamou and K.~Tobioka, \emph{{Constraining the charm
  Yukawa and Higgs-quark coupling universality}},
  \href{http://dx.doi.org/10.1103/PhysRevD.92.033016}{\emph{Phys. Rev.}
  {\bfseries D92} (2015) 033016},
  [\href{https://arxiv.org/abs/1503.00290}{{\ttfamily 1503.00290}}].

\bibitem{Carpenter:2016mwd}
L.~M. Carpenter, T.~Han, K.~Hendricks, Z.~Qian and N.~Zhou, \emph{{Higgs Boson
  Decay to Light Jets at the LHC}},
  \href{http://dx.doi.org/10.1103/PhysRevD.95.053003}{\emph{Phys. Rev.}
  {\bfseries D95} (2017) 053003},
  [\href{https://arxiv.org/abs/1611.05463}{{\ttfamily 1611.05463}}].

\bibitem{Han:2017yhy}
T.~Han and X.~Wang, \emph{{Radiative Decays of the Higgs Boson to a Pair of
  Fermions}},  \href{https://arxiv.org/abs/1704.00790}{{\ttfamily 1704.00790}}.

\bibitem{ATLAS:2017pfq}
{\scshape ATLAS} collaboration, T.~A. collaboration, \emph{{Constituent-level
  pile-up mitigation techniques in ATLAS}}, .

\bibitem{Berta:2014eza}
P.~Berta, M.~Spousta, D.~W. Miller and R.~Leitner, \emph{{Particle-level pileup
  subtraction for jets and jet shapes}},
  \href{http://dx.doi.org/10.1007/JHEP06(2014)092}{\emph{JHEP} {\bfseries 06}
  (2014) 092}, [\href{https://arxiv.org/abs/1403.3108}{{\ttfamily 1403.3108}}].

\bibitem{Bertolini:2014bba}
D.~Bertolini, P.~Harris, M.~Low and N.~Tran, \emph{{Pileup Per Particle
  Identification}},
  \href{http://dx.doi.org/10.1007/JHEP10(2014)059}{\emph{JHEP} {\bfseries 10}
  (2014) 059}, [\href{https://arxiv.org/abs/1407.6013}{{\ttfamily 1407.6013}}].

\bibitem{Krohn:2013lba}
D.~Krohn, M.~D. Schwartz, M.~Low and L.-T. Wang, \emph{{Jet Cleansing: Pileup
  Removal at High Luminosity}},
  \href{http://dx.doi.org/10.1103/PhysRevD.90.065020}{\emph{Phys. Rev.}
  {\bfseries D90} (2014) 065020},
  [\href{https://arxiv.org/abs/1309.4777}{{\ttfamily 1309.4777}}].

\bibitem{Cacciari:2014gra}
M.~Cacciari, G.~P. Salam and G.~Soyez, \emph{{SoftKiller, a particle-level
  pileup removal method}},
  \href{http://dx.doi.org/10.1140/epjc/s10052-015-3267-2}{\emph{Eur. Phys. J.}
  {\bfseries C75} (2015) 59},
  [\href{https://arxiv.org/abs/1407.0408}{{\ttfamily 1407.0408}}].

\bibitem{Cacciari:2007fd}
M.~Cacciari and G.~P. Salam, \emph{{Pileup subtraction using jet areas}},
  \href{http://dx.doi.org/10.1016/j.physletb.2007.09.077}{\emph{Phys. Lett.}
  {\bfseries B659} (2008) 119--126},
  [\href{https://arxiv.org/abs/0707.1378}{{\ttfamily 0707.1378}}].

\bibitem{CMS:2013wea}
{\scshape CMS} collaboration, C.~Collaboration, \emph{{Pileup Jet
  Identification}}, .

\bibitem{Aaboud:2017pou}
{\scshape ATLAS} collaboration, M.~Aaboud et~al., \emph{{Identification and
  rejection of pile-up jets at high pseudorapidity with the ATLAS detector}},
  \href{http://dx.doi.org/10.1140/epjc/s10052-017-5081-5,
  10.1140/epjc/s10052-017-5245-3}{\emph{Eur. Phys. J.} {\bfseries C77} (2017)
  580}, [\href{https://arxiv.org/abs/1705.02211}{{\ttfamily 1705.02211}}].

\bibitem{Aad:2015ina}
{\scshape ATLAS} collaboration, G.~Aad et~al., \emph{{Performance of pile-up
  mitigation techniques for jets in $pp$ collisions at $\sqrt{s}=8$ TeV using
  the ATLAS detector}},
  \href{http://dx.doi.org/10.1140/epjc/s10052-016-4395-z}{\emph{Eur. Phys. J.}
  {\bfseries C76} (2016) 581},
  [\href{https://arxiv.org/abs/1510.03823}{{\ttfamily 1510.03823}}].

\bibitem{Martinez:2018fwc}
J.~Martinez et~al., \emph{{Pileup mitigation at the Large Hadron Collider with
  Graph Neural Networks}},  \href{https://arxiv.org/abs/1810.07988}{{\ttfamily
  1810.07988}}.

\bibitem{Collaboration:2285584}
{ATLAS Collaboration}, \emph{{Technical Design Report for the Phase-II Upgrade
  of the ATLAS TDAQ System}},  Tech. Rep. CERN-LHCC-2017-020. ATLAS-TDR-029,
  CERN, Geneva, Sep, 2017.

\bibitem{Collaboration:2283192}
{CMS Collaboration}, \emph{{The Phase-2 Upgrade of the CMS L1 Trigger Interim
  Technical Design Report}},  Tech. Rep. CERN-LHCC-2017-013. CMS-TDR-017, CERN,
  Geneva, Sep, 2017.

\bibitem{Alwall:2014hca}
J.~Alwall, R.~Frederix, S.~Frixione, V.~Hirschi, F.~Maltoni, O.~Mattelaer
  et~al., \emph{{The automated computation of tree-level and next-to-leading
  order differential cross sections, and their matching to parton shower
  simulations}}, \href{http://dx.doi.org/10.1007/JHEP07(2014)079}{\emph{JHEP}
  {\bfseries 07} (2014) 079},
  [\href{https://arxiv.org/abs/1405.0301}{{\ttfamily 1405.0301}}].

\bibitem{Mangano:2002ea}
M.~L. Mangano, M.~Moretti, F.~Piccinini, R.~Pittau and A.~D. Polosa,
  \emph{{ALPGEN, a generator for hard multiparton processes in hadronic
  collisions}},
  \href{http://dx.doi.org/10.1088/1126-6708/2003/07/001}{\emph{JHEP} {\bfseries
  07} (2003) 001}, [\href{https://arxiv.org/abs/hep-ph/0206293}{{\ttfamily
  hep-ph/0206293}}].

\bibitem{Sjostrand:2006za}
T.~Sjostrand, S.~Mrenna and P.~Z. Skands, \emph{{PYTHIA 6.4 Physics and
  Manual}}, \href{http://dx.doi.org/10.1088/1126-6708/2006/05/026}{\emph{JHEP}
  {\bfseries 05} (2006) 026},
  [\href{https://arxiv.org/abs/hep-ph/0603175}{{\ttfamily hep-ph/0603175}}].

\bibitem{deFavereau:2013fsa}
{\scshape DELPHES 3} collaboration, J.~de~Favereau, C.~Delaere, P.~Demin,
  A.~Giammanco, V.~Lema{\^\i}tre, A.~Mertens et~al., \emph{{DELPHES 3, A
  modular framework for fast simulation of a generic collider experiment}},
  \href{http://dx.doi.org/10.1007/JHEP02(2014)057}{\emph{JHEP} {\bfseries 02}
  (2014) 057}, [\href{https://arxiv.org/abs/1307.6346}{{\ttfamily 1307.6346}}].

\bibitem{Aaboud:2016leb}
{\scshape ATLAS} collaboration, M.~Aaboud et~al., \emph{{Performance of the
  ATLAS Trigger System in 2015}},
  \href{http://dx.doi.org/10.1140/epjc/s10052-017-4852-3}{\emph{Eur. Phys. J.}
  {\bfseries C77} (2017) 317},
  [\href{https://arxiv.org/abs/1611.09661}{{\ttfamily 1611.09661}}].

\bibitem{Collaboration:2272264}
{CMS Collaboration}, \emph{{The Phase-2 Upgrade of the CMS Tracker}},  Tech.
  Rep. CERN-LHCC-2017-009. CMS-TDR-014, CERN, Geneva, Jun, 2017.

\bibitem{Collaboration:2283193}
{CMS Collaboration}, \emph{{The Phase-2 Upgrade of the CMS DAQ Interim
  Technical Design Report}},  Tech. Rep. CERN-LHCC-2017-014. CMS-TDR-018, CERN,
  Geneva, Sep, 2017.

\bibitem{Collaboration:2296612}
{CMS Collaboration}, \emph{{TECHNICAL PROPOSAL FOR A MIP TIMING DETECTOR IN THE
  CMS EXPERIMENT PHASE 2 UPGRADE}},  Tech. Rep. CERN-LHCC-2017-027. LHCC-P-009,
  CERN, Geneva, Dec, 2017.

\bibitem{Collaboration:2290829}
{ATLAS Collaboration}, \emph{{Expression of Interest: A High-Granularity Timing
  Detector for ATLAS Phase-2 Upgrade}},  Tech. Rep. ATL-COM-LARG-2017-049,
  CERN, Geneva, Oct, 2017.

\bibitem{Collaboration:2257755}
{ATLAS Collaboration}, \emph{{Technical Design Report for the ATLAS Inner
  Tracker Strip Detector}},  Tech. Rep. CERN-LHCC-2017-005. ATLAS-TDR-025,
  CERN, Geneva, Apr, 2017.

\bibitem{Collaboration:2285585}
{ATLAS Collaboration}, \emph{{Technical Design Report for the ATLAS Inner
  Tracker Pixel Detector}},  Tech. Rep. CERN-LHCC-2017-021. ATLAS-TDR-030,
  CERN, Geneva, Sep, 2017.

\bibitem{Sirunyan:2017ulk}
{\scshape CMS} collaboration, A.~M. Sirunyan3 et~al., \emph{{Particle-flow
  reconstruction and global event description with the CMS detector}},
  \href{http://dx.doi.org/10.1088/1748-0221/12/10/P10003}{\emph{JINST}
  {\bfseries 12} (2017) P10003},
  [\href{https://arxiv.org/abs/1706.04965}{{\ttfamily 1706.04965}}].

\bibitem{Aaboud:2017aca}
{\scshape ATLAS} collaboration, M.~Aaboud et~al., \emph{{Jet reconstruction and
  performance using particle flow with the ATLAS Detector}},
  \href{http://dx.doi.org/10.1140/epjc/s10052-017-5031-2}{\emph{Eur. Phys. J.}
  {\bfseries C77} (2017) 466},
  [\href{https://arxiv.org/abs/1703.10485}{{\ttfamily 1703.10485}}].

\bibitem{ATL-PHYS-PUB-2016-026}
{ATLAS Collaboration}, \emph{{Expected performance for an upgraded ATLAS
  detector at High-Luminosity LHC}},  Tech. Rep. ATL-PHYS-PUB-2016-026, CERN,
  Geneva, Oct, 2016.

\bibitem{ATL-PHYS-PUB-2017-013}
{ATLAS Collaboration}, \emph{{Optimisation and performance studies of the ATLAS
  $b$-tagging algorithms for the 2017-18 LHC run}},  Tech. Rep.
  ATL-PHYS-PUB-2017-013, CERN, Geneva, Jul, 2017.

\bibitem{Butler:2055167}
{CMS Collaboration}, \emph{{CMS Phase II Upgrade Scope Document}},  Tech. Rep.
  CERN-LHCC-2015-019. LHCC-G-165, CERN, Geneva, Sep, 2015.

\bibitem{CERN-LHCC-2015-020}
{ATLAS Collaboration}, \emph{{ATLAS Phase-II Upgrade Scoping Document}},  Tech.
  Rep. CERN-LHCC-2015-020. LHCC-G-166, CERN, Geneva, Sep, 2015.

\bibitem{Apollinari:2017cqg}
G.~Apollinari, O.~Brüning, T.~Nakamoto and L.~Rossi, \emph{{High Luminosity
  Large Hadron Collider HL-LHC}},
  \href{http://dx.doi.org/10.5170/CERN-2015-005.1}{\emph{CERN Yellow Report}
  (2015) 1--19}, [\href{https://arxiv.org/abs/1705.08830}{{\ttfamily
  1705.08830}}].

\bibitem{Read:2002hq}
A.~L. Read, \emph{{Presentation of search results: The CL(s) technique}},
  \href{http://dx.doi.org/10.1088/0954-3899/28/10/313}{\emph{J. Phys.}
  {\bfseries G28} (2002) 2693--2704}.

\bibitem{Chen_2016}
T.~Chen and C.~Guestrin, \emph{Xgboost},
  \href{http://dx.doi.org/10.1145/2939672.2939785}{\emph{Proceedings of the
  22nd ACM SIGKDD International Conference on Knowledge Discovery and Data
  Mining - KDD '16} (2016) }.

\bibitem{Louppe:2016ylz}
G.~Louppe, M.~Kagan and K.~Cranmer, \emph{{Learning to Pivot with Adversarial
  Networks}},  \href{https://arxiv.org/abs/1611.01046}{{\ttfamily 1611.01046}}.

\bibitem{Shimmin:2017mfk}
C.~Shimmin, P.~Sadowski, P.~Baldi, E.~Weik, D.~Whiteson, E.~Goul et~al.,
  \emph{{Decorrelated Jet Substructure Tagging using Adversarial Neural
  Networks}}, \href{http://dx.doi.org/10.1103/PhysRevD.96.074034}{\emph{Phys.
  Rev.} {\bfseries D96} (2017) 074034},
  [\href{https://arxiv.org/abs/1703.03507}{{\ttfamily 1703.03507}}].

\bibitem{Aguilar-Saavedra:2017rzt}
J.~A. Aguilar-Saavedra, J.~H. Collins and R.~K. Mishra, \emph{{A generic
  anti-QCD jet tagger}},
  \href{http://dx.doi.org/10.1007/JHEP11(2017)163}{\emph{JHEP} {\bfseries 11}
  (2017) 163}, [\href{https://arxiv.org/abs/1709.01087}{{\ttfamily
  1709.01087}}].

\bibitem{Stevens:2013dya}
J.~Stevens and M.~Williams, \emph{{uBoost: A boosting method for producing
  uniform selection efficiencies from multivariate classifiers}},
  \href{http://dx.doi.org/10.1088/1748-0221/8/12/P12013}{\emph{JINST}
  {\bfseries 8} (2013) P12013},
  [\href{https://arxiv.org/abs/1305.7248}{{\ttfamily 1305.7248}}].

\bibitem{Dolen:2016kst}
J.~Dolen, P.~Harris, S.~Marzani, S.~Rappoccio and N.~Tran, \emph{{Thinking
  outside the ROCs: Designing Decorrelated Taggers (DDT) for jet
  substructure}}, \href{http://dx.doi.org/10.1007/JHEP05(2016)156}{\emph{JHEP}
  {\bfseries 05} (2016) 156},
  [\href{https://arxiv.org/abs/1603.00027}{{\ttfamily 1603.00027}}].

\bibitem{Moult:2017okx}
I.~Moult, B.~Nachman and D.~Neill, \emph{{Convolved Substructure: Analytically
  Decorrelating Jet Substructure Observables}},
  \href{http://dx.doi.org/10.1007/JHEP05(2018)002}{\emph{JHEP} {\bfseries 05}
  (2018) 002}, [\href{https://arxiv.org/abs/1710.06859}{{\ttfamily
  1710.06859}}].

\bibitem{ATL-PHYS-PUB-2018-014}
{ATLAS Collaboration}, \emph{{Performance of mass-decorrelated jet substructure
  observables for hadronic two-body decay tagging in ATLAS}},  Tech. Rep.
  ATL-PHYS-PUB-2018-014, CERN, Geneva, Jul, 2018.

\bibitem{1742-6596-1067-2-022009}
J.~Abelleira et~al., \emph{{High-Energy LHC design}}, {\emph{Journal of
  Physics: Conference Series} {\bfseries 1067} (2018) 022009}.

\end{thebibliography}\endgroup
\end{document}